\documentstyle[preprint,aps]{revtex}
\begin{document}
\draft

\title{Wavelet Analysis of Vortex Tubes in Experimental Turbulence}

\author{Hideaki Mouri}
\address{Meteorological Research Institute, Nagamine 1-1, Tsukuba 305-0052, Japan}

\author{Masanori Takaoka}
\address{Department of Mechanical Engineering, Doshisha University, Kyotanabe, Kyoto 610-0321, Japan}

\date{\today}
\maketitle

\begin{abstract}
This report proposes a new method to study vortex tubes in one-dimensional velocity data of experimental turbulence. Vortex tubes are detected as local maxima on the scale-space plot of wavelet transforms of the velocity data. Then it is possible to extract a typical velocity pattern. The result for data obtained in a wind tunnel is consistent with those of three-dimensional direct numerical simulations.
\end{abstract}


By using direct numerical simulations and bubble/cavitation experiments, it has been established that turbulence contains vortex tubes \cite{j93,tmi99,d91,sa97}. Regions of intense vorticity are organized into tubes. Their radii and lengths are, respectively, of the orders of the Kolmogorov length $\eta$ and the integral length $L$. Vortex tubes occupy a small fraction of the volume and are embedded in a background flow, which is random and of large scales.

Especially when the Reynolds number $Re_{\lambda}$ is high, effects of vortex tubes on the velocity field are of interest \cite{sa97}. The velocity signal at small scales is enhanced only in a fraction of the volume. This fine-structure intermittency is attributable to vortex tubes. With increasing the Reynolds number, the flatness factor of the velocity derivative becomes high. This means that turbulence becomes more intermittent.

However, in direct numerical simulations, the Reynolds number is always low, $Re_{\lambda} \lesssim 200$. For the higher Reynolds numbers, one inevitably resorts to standard experiments, where a measurement is made with a probe suspended in the flow, and merely a one-dimensional cut of the velocity field is obtained. To study vortex tubes in such one-dimensional velocity data, we propose a new method.

The experiments often deal with the velocity components in the mean-flow direction alone (hereafter, the streamwise velocity $u$), but the transverse velocity $v$ is more suited to detecting rotational flows such as those associated with vortex tubes \cite{n97,cg99,mtk99,mhk00}. We accordingly use this velocity component.

Our method is based on orthonormal wavelets, i.e., self-similar functions localized both in scale and space. There are several known families of wavelets \cite{mk95,f92}. With a wavelet family, wavelet transforms are computed from the signal. Since a wavelet function has a zero mean, the transform corresponds to a variation in the signal at a given scale and a given position. Moreover, since a wavelet family constitutes a complete orthonormal basis, the transforms retain the same information as the original data. This advantage makes interpretation of the results reliable, especially in statistical analyses \cite{yo91,m99}.

Suppose that a function $v(x)$ is sampled as a discrete signal $v[n] = v(n \delta x)$, where $\delta x$ is the sampling interval ($n = 0$ to 2$^N-1$). The wavelet transformation is written as
\begin{equation}
\label{equ1}
v[n] = \langle v \rangle + \sum^{N-1}_{j=0} \sum^{2^j-1}_{k=0} \hat{v}_{j,k} w_{j,k}[n].
\end{equation}
Here $\langle \cdot \rangle$ denotes the average. The integers $j$ and $k$ specify the scale of the wavelet function $w_{j,k}[n]$ and its position on the $x$-axis, respectively. The wavelet transform $\hat{v}_{j,k}$ is the inner product of $v[n]$ with the wavelet function $w_{j,k}[n]$ and represents a signal variation of the scale $\ell = 2^{N-j-1} \delta x$ around the position $x = 2^{N-j} k \delta x$ \cite{yo91,m99}. 

We systematically detect vortex tubes with different scales and strengths as local maxima on the scale-space plot of $\hat{v}^2_{j,k}/\langle \hat{v}_{j,k}^2 \rangle$. Here $\langle \hat{v}_{j,k}^2 \rangle$ is the second-order moment computed for each of the scales, corresponds to its mean energy, and reflects both tubes and the background flow. The circulation flow of a tube of the size $\ell$ at the spatial position $x$ is expected to enhance the value of $\hat{v}^2_{j,k}/\langle \hat{v}_{j,k}^2 \rangle$ at the scale $\ell$ around the position $x$ \cite{cg99,mtk99,f92}. If the value of $\hat{v}^2_{j,k}/\langle \hat{v}_{j,k}^2 \rangle$ at a certain scale-space position $(j, k)$ is greater than those at adjacent positions, which are indicated by shaded areas in Fig.~\ref{Fig1}, we consider that a tube exists at that position. 

%%%% Figure 1 %%%%

The present analysis is based on Haar's wavelets, each of which is a sharp pulse in space:
\begin{equation}
w_{j,k}[n] = \left\{ \begin{array}{lll}
                     -2^{-\frac{N-j}{2}} & \mbox{for $n = 2^{N-j}k$ to $2^{N-j}k+2^{N-j-1}-1$}     \\
                     +2^{-\frac{N-j}{2}} & \mbox{for $n = 2^{N-j}k+2^{N-j-1}$ to $2^{N-j}(k+1)-1$} \\
                     0             & \mbox{elsewhere.}
                    \end{array}
             \right.
\end{equation}
Because of poor localization in scale, Haar's wavelets were not favored in previous studies. However, they have advantages. First, contrasting to other orthonormal wavelets that oscillate many times in space, Haar's wavelets represent single oscillations. Since the transverse-velocity profile of a vortex tube has the same character (see below), Haar's wavelets work better in analyses of vortex tubes. With the other wavelets, we have analyzed our data. A tube is detected repeatedly at distances from the wavelet center. Second, a Haar's transform is analogous to a velocity increment $\delta v(\ell)= v(x+\ell)-v(x)$, the standard tool to analyze experimental turbulence \cite{m99}. They have the same sign. Statistical results for Haar's transforms can be interpreted as those for velocity increments.

The most familiar model for a vortex tube is a Burgers vortex. This is an axisymmetric steady flow in a strain field. In cylindrical coordinates, they are written as
\begin{equation}
\label{equ3}
  u_{\Theta} \propto \frac{2 \nu}{a r} 
             \left[ 1 - \exp \left( - \frac{a r^2}{4 \nu} \right) \right]
  \quad (a > 0),
\end{equation}
and
\begin{equation}
\label{equ4}
\left( u_r , u_{\Theta}, u_z \right) = 
\left( - \frac{1}{2} a r, 0, a z\right).
\end{equation}
Here $\nu$ is the kinematic viscosity. The above equation (\ref{equ3}) describes a rigid-body rotation for small radii, and a circulation decaying in radius for large radii. The velocity is maximal at $r$ = $r_0$ = $2.24 (\nu / a)^{1/2}$. Thus $r_0$ is regarded as the tube radius.

Suppose that the vortex tube penetrates the $(x,y)$-plane at the point $(0,\Delta)$. Here the $x$- and $y$-axes are, respectively, in the streamwise and transverse directions. If the direction of the tube axis is $(\theta, \varphi)$ in spherical coordinates, the streamwise ($u$) and transverse ($v$) components of the circulation flow $u_{\Theta}$ are
\begin{equation}
\label{equ5}
u = \frac{\Delta \cos \theta}{r} u_{\Theta} (r) \quad {\rm and} \quad
v = \frac{x \cos \theta}{r}          u_{\Theta} (r),
\end{equation}
with
\begin{equation}
\label{equ6}
r^2 = x^2           ( 1 - \sin ^2 \theta \cos ^2 \varphi ) +
      \Delta ^2 ( 1 - \sin ^2 \theta \sin ^2 \varphi ) +
      2x \Delta \sin ^2 \theta \sin \varphi \cos \varphi.
\end{equation}
Likewise, for the radial inflow $u_r$ of the strain field (\ref{equ4}), the streamwise and transverse components are
\begin{equation}
\label{equ7}
u = \frac
    {x ( 1 - \sin ^2 \theta \cos ^2 \varphi ) + \Delta \sin ^2 \theta \sin \varphi \cos \varphi }
    {r} 
    u_r (r) 
\end{equation}
and
\begin{equation}
\label{equ8}
v = - \frac
{x \sin ^2 \theta \sin \varphi \cos \varphi + \Delta ( 1 - \sin ^2 \theta \sin ^2 \varphi )}
{r} 
u_r (r).
\end{equation}
If a tube passes close to the probe ($\Delta \lesssim r_0$) and the tube is not heavily inclined ($\theta \simeq 0$), the transverse velocity is dominated by the circulation flow. This situation is important. If $\Delta \gg r_0$, the velocity signal of the tube is weak and diluted by the background flow. If $\theta \gg 0$, the contribution of the circulation flow is small. The transverse velocity is then dominated by the radial inflow, which does not have a specific scale. To such scale-free motions, our method to detect tubes is insensitive. 

The velocity profiles of vortex tubes with $\Delta \lesssim r_0$ and $\theta \simeq 0$ are nearly the same \cite{mtk99,b96}. However, as indicated by the first term of Eq. (\ref{equ6}), the radius $r_x$ of a tube observed along the streamwise direction is different from its true radius $r_0$ as $r_x \simeq r_0 / (1- \sin ^2 \theta \cos ^2 \varphi )^{1/2}$. Thus self-similar functions such as Haar's wavelets capture efficiently those tubes. Although there might be contamination from tubes with $\Delta \gg r_0$ or $\theta \gg 0$, such a drawback is inherent in any of experimental works.

Hereafter we analyze data obtained in a wind tunnel \cite{mhk00}. Its test section was 3 $\times$ 2 $\times$ 18 m in size. Turbulence was produced by placing a grid across the entrance to the test section. The grid was made of two layers of uniformly spaced rods, the axes of which were perpendicular to each other. The separation of the axes of the adjacent rods was 40 cm. The cross section of the rod was 6 $\times$ 6 cm. We simultaneously measured the streamwise ($U+u$) and transverse ($v$) velocities with a hotwire anemometer. The probe was positioned on the tunnel axis at 6 m downstream of the grid. The signal was low-pass filtered at 8 kHz and sampled digitally at 16 kHz. The entire length of the signal was 2 $\times$ 10$^7$ points.

The mean streamwise velocity $U$ is 8.70 m s$^{-1}$. The root-mean-square fluctuations $\langle u^2 \rangle ^{1/2}$ and $\langle v^2 \rangle ^{1/2}$ are 0.446 and 0.427 m s$^{-1}$, respectively. Since the turbulence level $\langle u^2 \rangle ^{1/2} / U$ is less than 10\%, we rely on the frozen-eddy hypothesis of Taylor, $\partial / \partial t = - U \partial / \partial x$, which yields the integral length $L$ of 17.2 cm, the Taylor microscale $\lambda$ of 0.858 cm, the Kolmogorov length $\eta$ of 0.0270 cm, and the microscale Reynolds number $Re_{\lambda}$ of 260.

The velocity signal is divided into 2400 segments of 2$^{13}$ points. To each segment, we apply the wavelet transformation ($N = 13$). Then statistics are computed over the segments. We present results only for the scale $\ell = 8 \eta$ ($j$ = 10), i.e., the smallest scale to which our method to detect vortex tubes is applicable. Results at the other scales are similar. 

Fig.~\ref{Fig2} shows probability density functions (PDFs) of Haar's transforms (solid lines) and velocity increments (dotted lines). Since the transverse-velocity PDFs are symmetric, those of the absolute values are shown. The agreement between the wavelet transforms and velocity increments is excellent. 

%%% Figure 2 %%%

By averaging signals centered at the position where $\hat{v}^2_{j,k}/\langle \hat{v}_{j,k}^2 \rangle$ is locally maximal, we extract typical patterns of vortex tubes in the streamwise ($u$) and transverse ($v$) velocities. The detection rate of the local maxima per the integral length $L$ is 3.37 at $\ell = 8 \eta$. Since a wavelet function is spatially extended, we determine in each case the center position so that the absolute value of  the velocity increment $\vert \delta v (\ell ) \vert$ is maximal. When the increment is negative, we invert the sign of the $v$ signal. The result is shown in Fig.~\ref{Fig3} (solid lines). The $u$ pattern is shown separately for $\delta u > 0$ and $\delta u \le 0$ (designated, respectively, as $u^+$ and $u^-$). We also show velocity profiles (\ref{equ5}) and (\ref{equ7}) of a Burgers vortex.  It is assumed that the tube center passes through the probe position ($\Delta = 0$) and the tube axis is perpendicular to the streamwise and transverse directions ($\theta = 0$). The tube radius $r_x$ is determined so as to reproduce the $v$ pattern.

%%% Figure 3 %%%

The $v$ pattern of grid turbulence is close to the profile of a Burgers vortex. Since the $v$ pattern is somewhat extended, there might be additional contributions from vortex tubes with $\Delta \gg r_0$ or $\theta \gg 0$ and vortex sheets. Previously, velocity patterns of tubes were studied by averaging for large values of wavelet transforms or velocity increments \cite{cg99,mtk99,mhk00}. The pattern shapes are close to those in Fig.~\ref{Fig3}, but are biased toward strong tubes. Our results based on local maxima are more representative of vortex tubes in turbulence. 

The $u^{\pm}$ patterns of grid turbulence appear to be dominated by the circulation flow $u_{\Theta}$ of a vortex tube \cite{b96}. There is no significant evidence for the presence of the radial inflow $u_r$. A vortex tube is not necessarily identical to a Burgers vortex. The same conclusion was obtained from direct numerical simulations \cite{j93}.

%%% Figure 4 %%%

The tube radius $r_x = 10 \eta$ observed at $\ell = 8 \eta$ serves as an upper limit for the intrinsic tube radius $r_0$, which should be several of the Kolmogorov length $\eta$. With an increase of the scale $\ell$, there is increasing importance of inclined vortex tubes. There also exist tubes with intrinsically large radii \cite{hoy97}. On the other hand, the amplitude of the $v$ pattern, corresponding to the circulation velocity of a typical tube, is of the order of $\langle v^2 \rangle ^{1/2}$. The radius of a tube and its circulation velocity estimated here are consistent with those obtained from direct numerical simulations \cite{j93,tmi99}.

The statistical fluctuation of the transverse velocity $v$ is shown in Fig.~\ref{Fig4}. Vortex tubes are embedded in a large-scale background flow. Hence, from the velocity signals around the local maxima of $\hat{v}^2_{j,k}/\langle \hat{v}_{j,k}^2 \rangle$, the expansion formula (\ref{equ1}) had been used to remove motions with scales $\ell > 8 \eta$ $(j < 10)$. The resultant fluctuation, which mainly reflects differences of tube parameters, is comparable to that in a direct numerical simulation \cite{tmi99}.

%%% Figure 5 %%%

The PDFs of wavelet transforms and velocity increments for the transverse velocity $v$ in Fig.~\ref{Fig2}(b) have long tails \cite{sa97,mhk00,m99}. It is possible to study this intermittency phenomenon at each scale by comparing a PDF of wavelet transforms at local maxima of $\hat{v}^2_{j,k}/\langle \hat{v}_{j,k}^2 \rangle$ with a PDF of the whole transforms as in Fig.~\ref{Fig5} (solid lines). The PDF of the local-maximum transforms has a peak at $\vert \hat{v}_{j,k} \vert \simeq \langle \hat{v}^2_{j,k} \rangle ^{1/2}$ and a long tail toward large magnitudes. If the condition for the local maximum is relaxed so as to ignore transforms at the smaller scale $j+1$, the result reproduces perfectly the tail of the whole-transform PDF (dotted line). Thus vortex tubes account for the long tail of the PDF. The situation is the same in velocity increments, since Haar's transforms are statistically equivalent to velocity increments. However, only 10\% of the scale energy is shared by the above subset of wavelet transforms that reproduce the tail of the whole-transform PDF. The background flow is energetically predominant \cite{b96}.

Finally, we give additional comments. First, our method to study vortex tubes is crude. They have been studied with three-dimensional direct numerical simulations in more reliable ways \cite{j93,tmi99}. Nevertheless, our method is useful at high Reynolds numbers, where only one-dimensional experimental data are available. Second, although our method is based on wavelet transforms for the transverse velocity alone, those for the streamwise velocity are also useful. For example, they could constrain the local strain field (see Fig.~\ref{Fig2}(a)). Third, our method is intermediate between those of our previous works. We proposed statistical measures based on orthonormal wavelets \cite{m99}. Although these measures are not relevant directly to tubes, they are robust with respect to the choice of the wavelets and thus characterize rigorously scale-space structures of a velocity signal. We also proposed a method to detect tubes by using the transverse-velocity profile of a model tube as a nonorthonormal wavelet \cite{mtk99}. Although a nonorthonormal wavelet transformation is redundant, this method provides high-resolution estimates of the size and spatial position of a tube. We hope that the application of these methods would improve knowledge of small-scale structures of turbulence.

%%% References %%%

%%%% Figure Captions %%%%

\begin{figure}
\caption{
Schematic representation of wavelet transforms on a scale-space plot. If the value of $\hat{v}^2_{j,k} / \langle \hat{v}^2_{j,k} \rangle$ at the position $(j,k)$ exceeds those at the adjacent positions (shaded areas), we consider that $\hat{v}^2_{j,k} / \langle \hat{v}^2_{j,k} \rangle$ is locally maximal at $(j,k)$.}
\label{Fig1}
\end{figure}

\begin{figure}
\caption{
PDFs of wavelet transforms (solid lines) and velocity increments (dotted lines) at $\ell = 8 \eta$ for the streamwise (a) and transverse (b) velocities. The abscissa is normalized by the root-mean-square value: $\langle \hat{u}_{j,k}^2 \rangle ^{1/2}$ = 0.124, $\langle \delta u^2 \rangle ^{1/2}$ = 0.0950, $\langle \hat{v}_{j,k}^2 \rangle ^{1/2}$ = 0.169, and $\langle \delta v ^2 \rangle ^{1/2}$ = 0.132 m s$^{-1}$. We also show Gaussian distributions with zero means and unity standard deviations.}
\label{Fig2}
\end{figure}

\begin{figure}
\caption{
Conditional averages of the streamwise ($u$) and transverse ($v$) velocities for local maxima of $\hat{v}^2_{j,k}/\langle \hat{v}_{j,k}^2 \rangle$ at $\ell = 8 \eta$ (solid lines). The abscissa is the spatial position $x$ normalized by the Kolmogorov length $\eta$. We show the streamwise velocity separately for $\delta u > 0$ $(u^+)$ and $\delta u \le 0$ $(u^-)$. The dotted lines represent profiles of a Burgers vortex.}
\label{Fig3}
\end{figure}

\begin{figure}
\caption{
Conditional average of the transverse velocity $v$ for local maxima of $\hat{v}^2_{j,k}/\langle \hat{v}_{j,k}^2 \rangle$ at $\ell = 8 \eta$. The abscissa is the spatial position $x$ normalized by the Kolmogorov length $\eta$. The error bars denote the statistical fluctuation ($\pm 1 \sigma$), which has been computed after we had removed motions with scales $\ell > 8 \eta$.}
\label{Fig4}
\end{figure}

\begin{figure}
\caption{
PDFs of wavelet transforms of the transverse velocity $\vert \hat{v}_{j,k} \vert$ at $\ell = 8 \eta$. The abscissa is normalized by the root mean square $\langle \hat{v}_{j,k}^2 \rangle ^{1/2}$ of the whole transforms at the scale. We show PDFs for local maxima of $\hat{v}^2_{j,k}/\langle \hat{v}_{j,k}^2 \rangle$ and for the whole transforms (solid lines), normalized by the number of the whole transforms. The fraction of the scale energy shared by the local-maximum transforms is 5.24\%. We also show a PDF for the local maxima determined without the smaller scale $j+1$ (dotted line). Their energy fraction is 9.56\%.}
\label{Fig5}
\end{figure}


\begin{references}
\bibitem{j93} J. Jim\'enez, A.A. Wray, P.G. Saffman, and R.S. Rogallo, J. Fluid Mech. {\bf 255}, 65 (1993); J. Jim\'enez and A.A. Wray, {\it ibid.} {\bf 373}, 255 (1998).
\bibitem{tmi99} M. Tanahashi, T. Miyauchi, and J. Ikeda, in {\it IUTAM Symposium on Simulation and Identification of Organized Structures in Flows}, edited by J.N. S$\o$rensen, E.J. Hopfinger, and N. Aubry (Kluwer, Dordrecht, 1999), p. 131.
\bibitem{d91} S. Douady, Y. Couder, and M.E. Brachet, Phys. Rev. Lett. {\bf 67}, 983 (1991); A. La Porta, G.A. Voth, F. Moisy, and E. Bodenschatz, Phys. Fluids {\bf 12}, 1485 (2000).
\bibitem{sa97} K.R. Sreenivasan and R.A. Antonia, Annu. Rev. Fluid Mech. {\bf 29}, 435 (1997).
\bibitem{n97} A. Noullez, G. Wallace, W. Lempert, R.B. Miles, and U. Frisch, J. Fluid Mech. {\bf 339}, 287 (1997).
\bibitem{cg99} R. Camussi and G. Guj, Phys. Fluids {\bf 11}, 423 (1999).
\bibitem{mtk99} H. Mouri, M. Takaoka, and H. Kubotani, Phys. Lett. A {\bf 261}, 82 (1999).
\bibitem{mhk00} H. Mouri, A. Hori, and Y. Kawashima, Phys. Lett. A {\bf 276}, 115 (2000).
\bibitem{mk95} H. Mouri and H. Kubotani, Phys. Lett. A {\bf 201}, 53 (1995).
\bibitem{f92} M. Farge, Annu. Rev. Fluid Mech. {\bf 24}, 395 (1992).
\bibitem{yo91} M. Yamada and K. Ohkitani, Prog. Theor. Phys. {\bf 86}, 799 (1991); C. Meneveau, J. Fluid Mech. {\bf 232}, 469 (1991).
\bibitem{m99} H. Mouri, H. Kubotani, T. Fujitani, H. Niino, and M. Takaoka, J. Fluid Mech. {\bf 389}, 229 (1999).
\bibitem{b96} F. Belin, J. Maurer, P. Tabeling, and H. Willaime, J. Phys. (France) II {\bf 6}, 573 (1996).
\bibitem{hoy97} I. Hosokawa, S. Oide, and K. Yamamoto, J. Phys. Soc. Jpn. {\bf 66}, 2961 (1997).
\end{references}
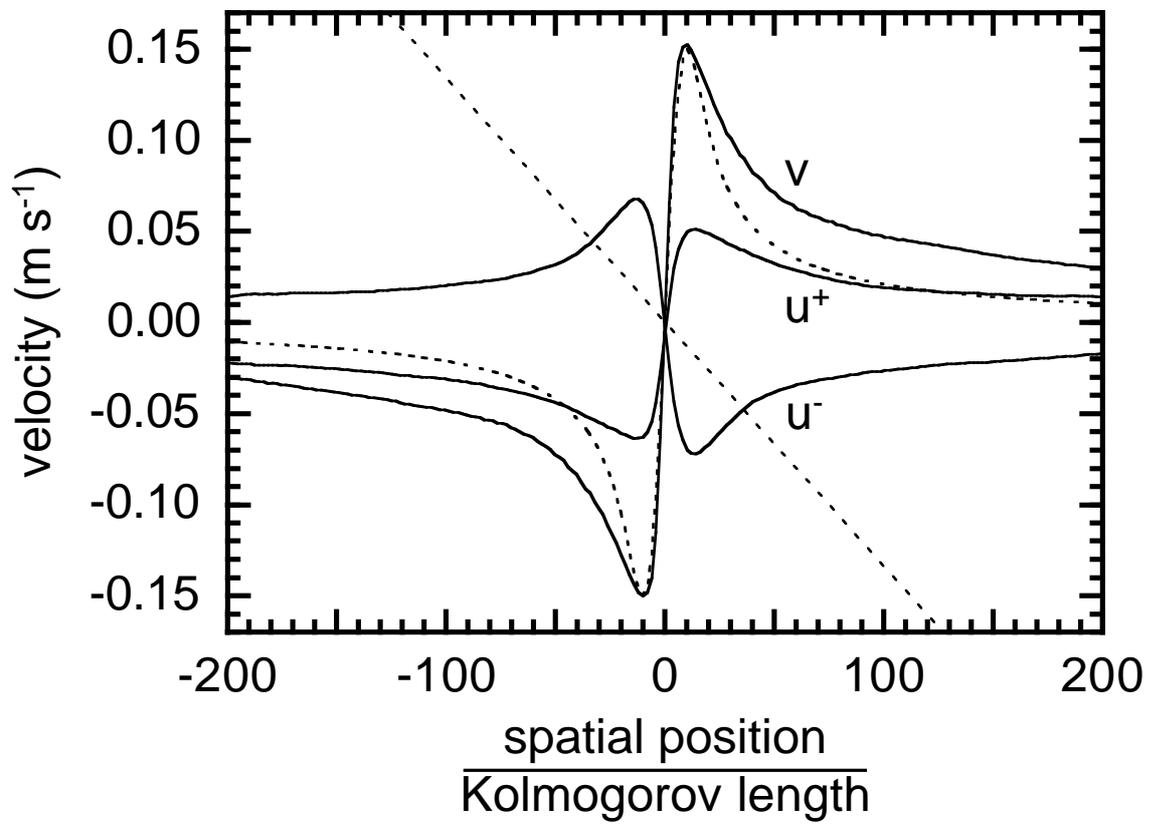
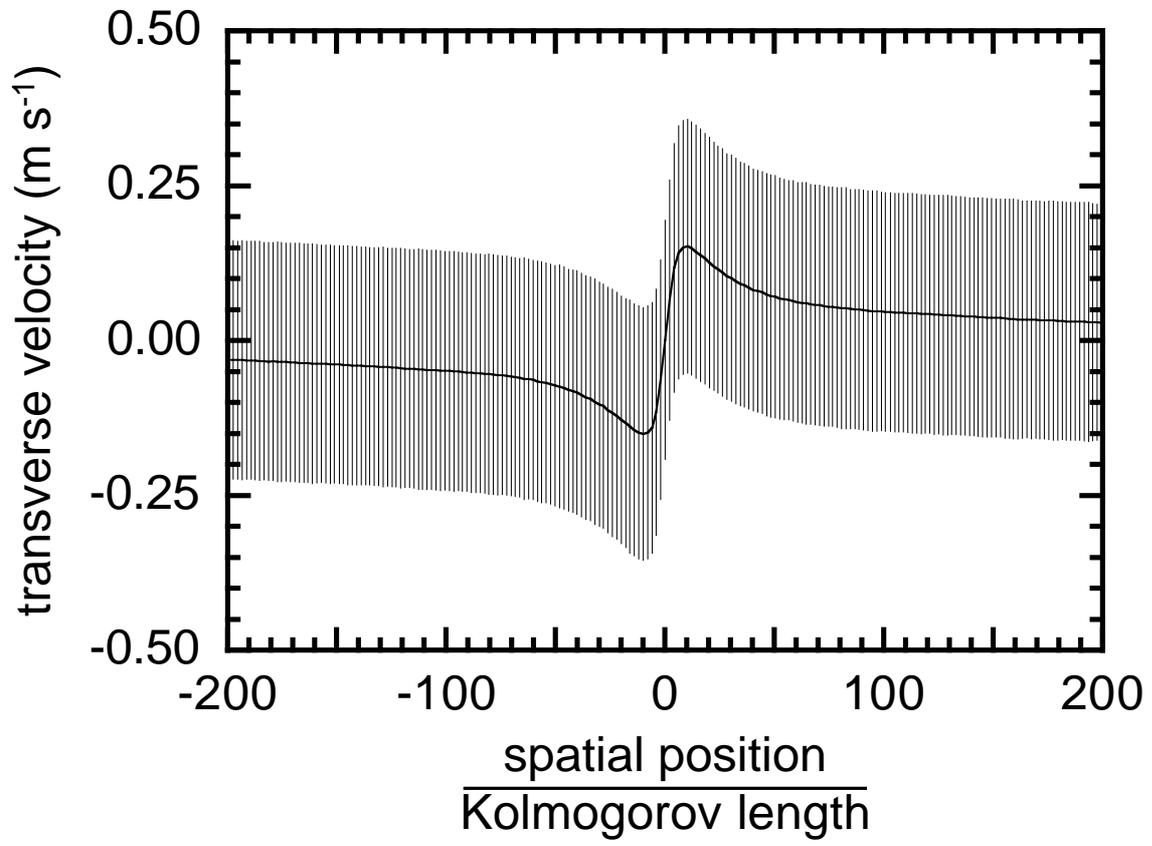
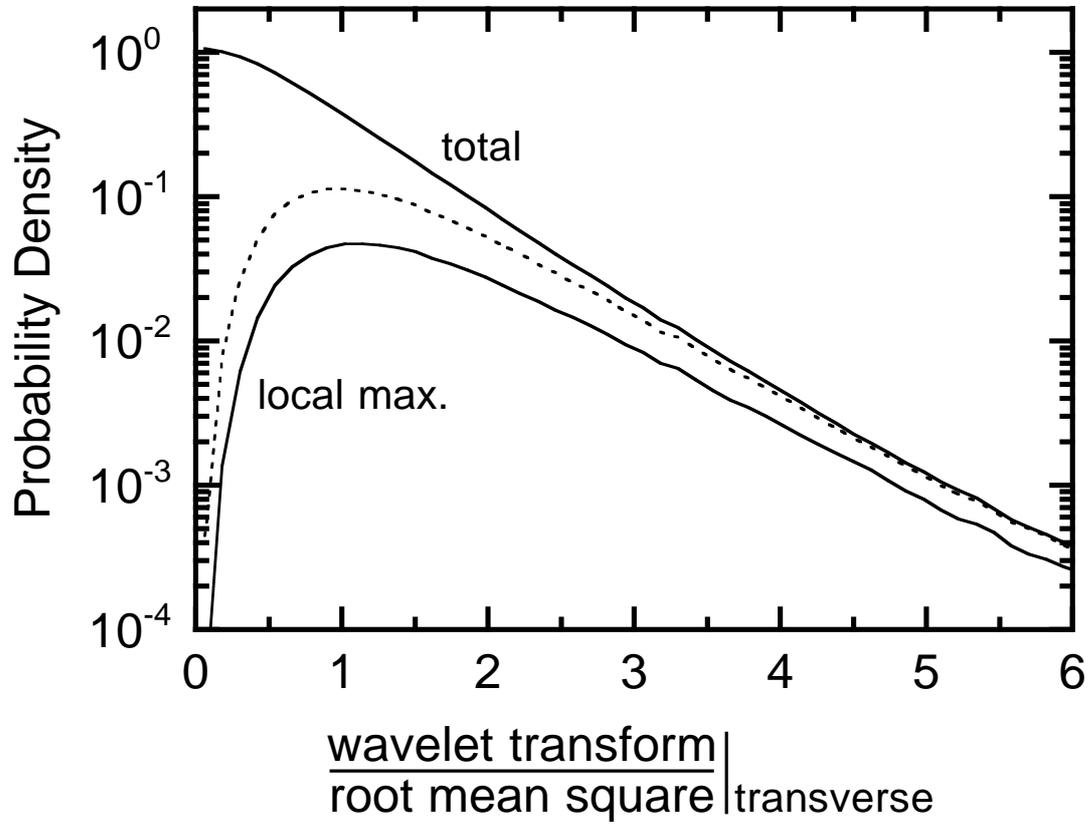
\end{document}